\documentclass[aps,prl,twocolumn,letterpaper]{revtex4}
\ifx\pdftexversion\undefined
  \usepackage[dvips]{graphicx}
\else
  \usepackage[pdftex]{graphicx}
\fi

\def\ie{{\em i.e.\ }}

\def\d{{\rm d}}

\def\dd#1#2{\frac{\d #1}{\d #2}}
 
\def\rmmat#1{{\hbox{\rm #1}}}
\def\rmscr#1{\rmmat{\scriptsize #1}}

\begin{document}

\title{The Long-Term Future of Space Travel}
\author{Jeremy S. Heyl}
\affiliation{Department of Physics and Astronomy, 
University of British Columbia \\ 
6224 Agricultural Road, Vancouver, British Columbia, Canada, V6T 1Z1;
Canada Research Chair}
\date{\today}

\begin{abstract}
The fact that we apparently live in an accelerating universe places
limitations on where humans might visit.  If the current energy
density of the universe is dominated by a cosmological constant, a
rocket could reach a galaxy observed today at a redshift of 1.7 on a
one-way journey or merely 0.65 on a round trip.  Unfortunately these
maximal trips are impractical as they require an infinite proper time
to traverse.  However, calculating the rocket trajectory in detail
shows that a rocketeer could nearly reach such galaxies within a
lifetime (a long lifetime admittedly -- about 100 years).  For less
negative values of $w$ the maximal redshift increases becoming
infinite for $w\geq -1/3$.  
\end{abstract} 


\maketitle

\section{Introduction}

It is happy coincidence that product of the lifespan of a human being
and the surface gravity of the Earth $\tau g/c$ is much larger than
unity; this means that at least special relativity presents no
impediment to interstellar travel \cite[\protect{\em e.g.}][]{Misn73}.
However, the recent discovery that the expansion of the universe is
accelerating \cite{1998AJ....116.1009R,2003ApJS..148..175S} presents a
new set of challenges for long-distance travel.  Several authors have
pointed out that because the universe is accelerating, we are now
seeing as much of the universe as we will ever see
\cite{2000ApJ...531...22K} and that the long-term future of extra
Galactic astronomy is bleak with only the handful of galaxies bound to
the Local Group ultimately being observable
\cite{2002PhRvD..65d7301L}.

The presence of a horizon that limits what we can observe in the
future also limits where we can visit.  However, unlike light humans
age as they travel and must be accelerated not too roughly to reach
relativistic velocities (and it turns out to maintain them as well as
we shall see).  Several authors have examined the related question of
how an accelerating expansion limits the future of computation \cite{
Kras04,2003PhLB..566....1B}.  However, computers can
survive larger accelerations than we can and possibly can operate over
a longer proper time.

In first section I shall derive the equations that describe
accelerated motion in an expanding universe.  The universe is
apparently entering an epoch of de Sitter expansion and solutions for
this spacetime may be solved analytically. Furthermore, I place
interesting analytic limits on the proper time required to travel
cosmological distances in more general spacetimes.  The second section
describes the results of numerical calculations of accelerating
trajectories through universes like our own and specifically focusses
on paths that will take approximately a human lifetime to traverse.

\section{Calculations}
\label{sec:calculations}

The metric of a homogeneous, isotropic, flat universe may be given by
\begin{equation}
ds^2 = g_{\alpha\beta} dx^\alpha dy^\beta = a(\tau)^2 \left [ d \tau^2 -
  dx^2 - dy^2 - dz^2 \right ] 
\label{eq:1}
\end{equation}
where $d t = a(\tau) d \tau$ and $c=1$. The symbol $\tau$ represents
the conformal time while $t$ is the time measured by a comoving
observer.  I will generally suppress the functional dependence of
$a(\tau)$, writing $a$.

A rocket traveller must have a four-velocity ($u^\alpha$) that satisfies
\begin{equation}
1 = u^\alpha u^\beta g_{\alpha\beta}.
\label{eq:2}
\end{equation}
Let us assume that the rocket travels in the $x-$direction yielding
\begin{equation}
\dd{\tau}{s}= u^0 = \frac{\cosh \chi }{a} ~\rmmat{and}~
\dd{x}{s} = u^1=\frac{\sinh\chi}{a}
\label{eq:3}
\end{equation}
that automatically satisfies Eq.~(\ref{eq:2}).  The symbol $\chi$ denotes
the rapidity of the rocket that a coincident,
comoving observer would measured. 

The acceleration of the rocket measured by the rocketeer is given by
\begin{eqnarray}
\xi^\alpha &=& u^{\alpha}_{;\beta} u^\beta \\
 &=& \dd{u^\alpha}{x^\beta} u^\beta +
\Gamma^\alpha_{\gamma\beta} u^\gamma u^\beta = \dd{u^\alpha}{s} +
\Gamma^\alpha_{\gamma\beta} u^\gamma u^\beta 
\\
&=& 
\left ( \dd{\chi}{s} + \dd{a}{\tau} \frac{1}{a} u^1 \right )\left [
\begin{array}{c} 
u^1   \\
u^0 \\
0 \\
0 
\end{array}
\right ]
\end{eqnarray}
and
\begin{equation}
\xi^2 = -\left ( \dd{\chi}{s} + \dd{a}{\tau} \frac{1}{a} u^1 \right )^2 = 
-\left ( \dd{\chi}{s} + H \sinh \chi \right )^2
\label{eq:4}
\end{equation}
where $H=\dd{a}{\tau} a^{-2}$ is the Hubble parameter.  Let's fix the
amplitude of the proper acceleration $\xi^{\alpha}$ to be equal to the
gravitational acceleration at the surface of the Earth, $g$, and find
an equation describing how the rapidity changes with proper time.

In the Friedmann-Robertson-Walker spacetime we get
\begin{equation}
\dd{\chi}{s} = g - H \sinh\chi.
\label{eq:5}
\end{equation}
The second term is a friction term due to the expansion of the
universe.  If $a(\tau)$ is constant (\ie $H=0$), we recover Minkowsky
spacetime and the usual result that $\chi$ is the product of the
proper time of the rocket and its acceleration.  If we equate these
two terms we find the maximum value of the rapidity for a fixed proper
acceleration
\begin{equation}
\sinh \chi = \beta \gamma = \frac{g}{H} \approx 1.4 \times 10^{10}.
\label{eq:6}
\end{equation}
An object that exceeds this rapidity will experience an acceleration 
that exceeds $g$, so Eq.~(\ref{eq:6}) is a practical speed limit for
humans in an expanding universe.

Let's specialize to de Sitter space where $H$ is constant to get
\begin{eqnarray}
\chi(s) &=& 2 \tanh^{-1} \left \{ \frac {\tanh \left[ \frac{1}{2}
    \left ( s - C \right ) g' \right ] g' - H}{g} \right \}
\label{eq:7} \\
\tau(\chi) &=& \frac{\sinh \chi}{g}, ~~~ x(\chi) = \frac{\cosh\chi - 1}{g} \\
t(\chi) &=& -\frac{1}{H} \ln \left (1 - \frac{H}{g} \sinh \chi \right
) \\
l(\chi) &=& \frac{1}{H} \left [ 2 \frac{g}{g'} \tanh^{-1} \left ( \frac{g}{g'} \tanh
\frac{\chi}{2} + \frac{H}{g'} \right ) - \chi \right ]
\end{eqnarray}
where $g'=\sqrt{g^2+H^2} \approx g$, $C=-2 (g')^{-1}\tanh^{-1} (H/g')$
and $l$ is the proper distance ($dl = a d x$)
along the path.  We have taken $\chi=\tau=x=t=l=0$ at $s=0$. Because $g\gg H$ we have
\begin{equation}
\chi(s) = g ( s - C )
\label{eq:8}\end{equation}
when $ H (g')^{-2} \ll (s - C) \ll g^{-1}\ln(g/H)$. 
For $g=980$~m~s$^{-2}$, we have $s-C \gg 3$~ms and the second term in
the numerator of Eq.~\ref{eq:7} within the braces dominates.  The linear increase of
$\chi$ with proper time stops when the condition given by
Eq.~(\ref{eq:6}) is approached; that is, when $s \approx g^{-1} \ln
(g/H) \approx 23$~years.

Even if $H$ varies with time, we can find a solution to
Eqs.~(\ref{eq:3}) and~(\ref{eq:5}) for $g=0$: $a \sinh \chi = a_0$  \cite{Pebb93}
where $a_0$ is a constant of integration.  We see that the value of
$\beta \gamma a$ is constant along a geodesic of a massive particle in
the metric given by Eq.~(\ref{eq:1}).  We find that a
particle travelling along a geodesic with finite rapidity today only
has an infinite rapidity as $a \rightarrow 0$. 

Let's return to the general case with $g\neq 0$ and try to obtain a
bound on the proper time required to travel between two scale factors.
After the initial increase in the rapidity until $s \approx g^{-1} \ln
(g/H_0) \approx 23$~years, the rapidity saturates so we have
\begin{equation}
\dd{\tau}{s} \approx \dd{x}{s} \approx \frac{g}{H a}.
\label{eq:16}
\end{equation}
In a general Robertson-Walker spacetime, the Hubble parameter changes
with time, but it is reasonable to assume that Eq.~(\ref{eq:16}) still
holds as long as $H$ does not change drastically over the timescale
$1/H$.  In this case we find that the elapsed proper time of the
rocketeer is simply related to the ratio of the scale factor at the
beginning of the friction dominated era to current scale factor 
\begin{equation}
s_f - s_i \approx
 \frac{1}{g} \ln \left ( \frac{a_f}{a_i} \right ).
\label{eq:17}
\end{equation}
The preceding equation provides a crude estimate of the proper time.  

If the Hubble parameter is increasing with time ($w<-1$)
Eq.~(\ref{eq:17}) provides an approximate upper bound on the proper
time and conversely it provides an approximate lower bound if $w>-1$
and the Hubble parameter decreases with time.  The error in this
estimate is most dramatic for $w<-1$.  In this case, the curvature of
the universe diverges ($a\rightarrow \infty$) within a finite proper
time along geodesics \cite{2003PhRvL..91g1301C}.
Because geodesics are paths between events that
maximize the proper time, we find in general for $w<-1$ the rocketeer
reaches $a\rightarrow \infty$ within a finite proper time, in fact
within a shorter time than if she stayed at home.

On the other hand if the Hubble parameter is constant or decreasing
with time ($w\geq -1$), the condition $\sinh \chi = g/H$ is only
approached but never reached within a finite proper time, so we have
from Eq.~(\ref{eq:3})
\begin{eqnarray}
a \dd{\tau}{s} &<& \sqrt{1 + \frac{g^2}{H^2}} \\
\dd{a}{\tau} \frac{1}{a} \dd{\tau}{s} &<& \sqrt{H^2 + g^2} \\
\frac{1}{a\sqrt{H^2+g^2}} \d a &<& \d s .
\end{eqnarray}
Integrating this for a particular value of $w=P/\rho$ so that $H^2
\propto a^{-(3w+3)}$, the equation of
state, yields
\begin{equation}
\Delta s > \left \{ \begin{array}{ll} 
(g')^{-1} \ln \left (\frac{a_f}{a_i} \right ) & \rmmat{if } w=-1 \\
\left . \frac{2}{3w+3} g^{-1} \tanh^{-1}\left ( \frac{g}{g'} \right )
\right |_{a_i}^{a_f} &
\rmmat{if } w>-1. \\
\end{array} \right . 
\end{equation}
As a reminder $(g')^2=H_0^2 a^{-(3w+3)}+g^2$.  If we take $a_f
\rightarrow \infty$ we find that both expressions diverge, so for
$w>-1$ the rocketeer requires an infinite proper time to reach
$a=\infty$.

\section{Numerical Results}
\label{sec:numerical-results}

\begin{figure}
\centerline{\includegraphics[height=9cm]{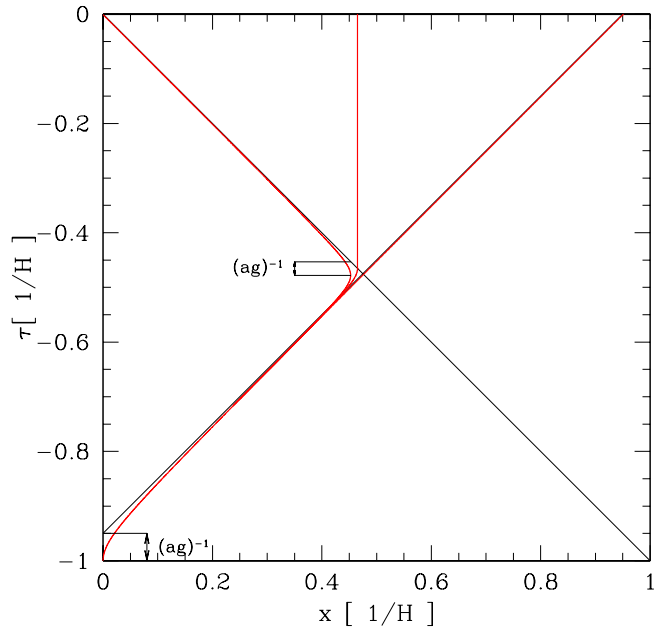} }
\smallskip
\caption{Three possible paths (bold curves) for the rocketeer for
  deSitter space and the light cones that these curves asymptotically
  approach.
  On the lowermost path, the rocket is constantly accelerating and
  sends a signal back to Earth that arrives on Earth after an infinite
  time.  The middle path shows a rocket that decelerates and stops
  relative to the comoving material and sends a signal back to Earth
  and the uppermost path shows a rocketeer that returns to Earth
  albeit after an infinite proper time.}
\label{fig:path}
\end{figure}

Before delving into the calculational details for our universe with
$\Omega_1 \approx 0.27, w_1=0$ and $\Omega_2 \approx 0.73, w_2 \approx
-1$ and nearly flat spatial hypersurfaces
\cite{1998AJ....116.1009R,2003ApJS..148..175S}, let's outline three
possible trajectories for the rocket.  Fig.~\ref{fig:path} depicts
these choices for the trips of infinite elapsed proper time if $-1/3 >
w_2 \geq -1$.  Because the rocket's path asymptotically approaches a
null geodesic it is straightforward to calculate the maximum comoving
distance that our rocket can travel.  If it returns or ends at rest
with respect to the locally comoving material at a position where it
still can communicate with Earth we have $x_\rmscr{max} = 1/(2H)$
(with an error on the order of $g^{-1}$ ).  If the rocket continually
accelerates it will reach $x_\rmscr{max}=1/H$ after an infinite proper
time has elapsed.  If universe stops accelerating in the future, the
rocket can travel arbitrarily far in an arbitrarily long proper time.
Conversely, if $w_2<-1$, a comoving observer will find that the
universe only will exist for a finite proper time and it is no
different for the rocketeer.

Regardless of the details of the cosmology, on the outbound trip the
Earth can send messages and in principle energy for a time
$(1+a_f/a_i)/g$ where $a_f/a_i$ is the ratio of the initial scale
factor at launch to the final scale factor during the deceleration.
During the return trip the rocket can receive messages from Earth
emitted between a time $(1+a_f/a_i)/g$ after launch until the rocket
returns.  Conversely, all the messages sent from the rocket to Earth
during the outbound phase arrive between the launch and a time before
$(a_r/a_f)/g$ of the rocket's return where $a_r$ is the scale factor
of the universe when the rocket returns.  Messages sent during the
return trip arrive within a time $(a_r/a_f)/g$ of the rocket's return.

In practice because $g \gg H$ the three possibilities result in a
similar comoving distance travelled (at the moment of last contact) --
this difference in the comoving distance that the rocketeer travels
where it can make a final communication with Earth is only $(ag)^{-1}
\ll\ H^{-1}$ However, none of these paths are realistic because they
take an infinite proper time to traverse. It turns out that paths that
travel 90\% or even 99\% of the maximal comoving distance take about
only 100 years to traverse, so paths displayed in Fig.~\ref{fig:path}
are interesting to study even for realistic journeys.

The maximum comoving distance that one can reach is simply the
conformal time remaining until $a\rightarrow \infty$ ($1.12339/H_0$
for $w=-1, \Omega_{M,0}=0.27$ and $\Omega_{\Lambda,0}=0.73$) and half
that comoving distance for the round-trip journey.   What is more
interesting to know is the current age and redshift of objects that we
observe today at the particular comoving distances, so one could say
``Let's go to that galaxy'' and know whether it is possible.
\begin{figure}
\centering
\centerline{\includegraphics[height=9cm]{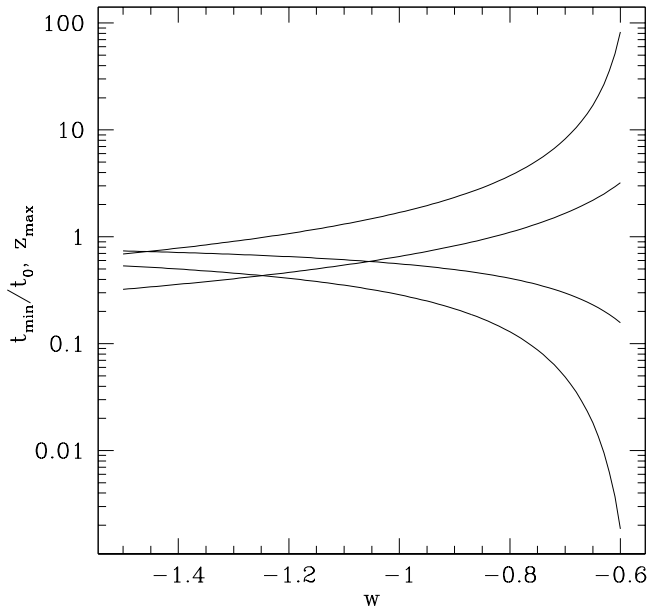}}
\smallskip
\caption{ The currently observed redshift (upper set of curves) and
age of galaxies (lower set of curves) located at the maximal comoving
distances that can be achieved on a one-way (outer curves) and
round-trip journeys (inner curves).
}
\label{fig:max}
\end{figure}

Fig.~\ref{fig:max} shows the current redshift of objects observed
today at the maximal comoving distance that can be achieved during
one-way and round-trip journeys.  Essentially for the favoured value
of $w=-1$, our rocketeer could visit a galaxy at a redshift of 0.65 if
she intended to return to the Milky Way (or report her results) or 1.7
if she didn't intend to come back.  As the value of $w$ approaches
$-1/3$, she could travel to further and further distances because the
acceleration of the universe is weaker. At $w=-0.6$ we have $z=3.2$
(round-trip) or $z=83$ (one-way).  An important point to keep in mind
is that these ``one-way'' trips are literally one-way.  At the end of
the rocket's journey, the rocket lies outside the asymptotic
past-light cone of the Milky Way (to the right of the diagonal line in
Fig.~\ref{fig:path}).

\begin{figure}
\centerline{\includegraphics[height=9cm]{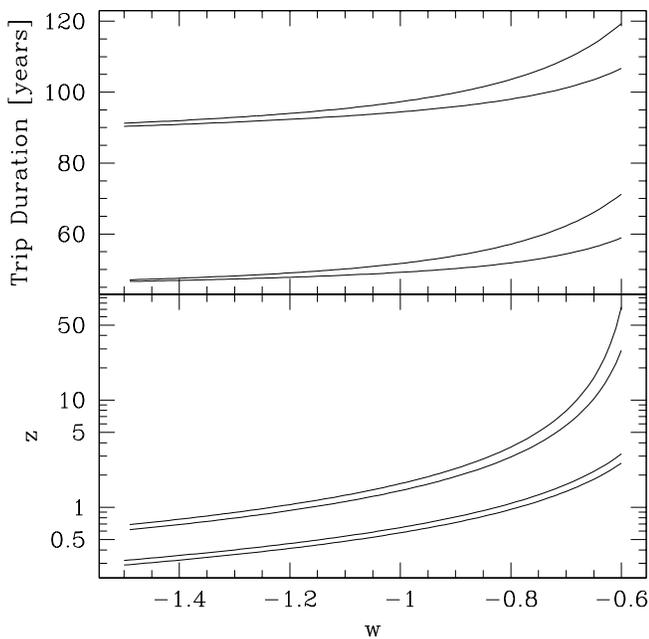}}
\smallskip
\caption{ The upper panel shows the duration of the journey according
 to the rocketeer The upper curves are for the round-trip journey
 lasting 99\% and 90\% of the remaining conformal time.  The lower
 curves show the currently observed value of redshift at the far end
 of the journey.  The upper set is for one-way trips, and the lower
 set is for round trips.  }
\label{fig:duration}
\end{figure}
To be more realistic we would like to see how long would a journey
last that almost reaches the maximum distance (reaching the maximum
distance takes an infinite proper time).  Specifically, we use 90\%
and 99\% of the maximal comoving distance that can be traversed on
one-way and round-trip journeys.  Typically, round-trips take
approximately 100 years and one-way trips take a bit more than half as
long, but the one-way trips of course reach much higher comoving
distances.  The reason for this is that the distance travelled by the
rocket increases exponentially until the rocket is travelling a
cosmological distance each year approximately; therefore, only the
portion of the journey with $s \gtrsim g^{-1} \ln (g/H_0) \approx
23$~years contributes significantly.  The round-trip journeys
typically accelerate for only slightly more than 23~years ($23.3$ and
$23.5$ for the case of $w=-1$, whereas the
one-way journeys accelerate for a few more years before slowing
down, so they travel twice the comoving distance.

The journeys that travel to 99\% of the maximum distance only last
slightly longer in the frame of the rocket than journeys that reach
only 90\% of the maximal distance.  However, the elapsed time on Earth
(or whatever remains of the Earth) can differ by up to an order of
magnitude for $w \sim -0.6$ or at least a factor of two for $w>-1$.
For $w=-1$ the round-trip lasts about 36 Gyr in the frame of the Milky
Way for the shorter journey and 72 Gyr for the longer journey.  The
vast majority of the elapsed comoving time is during the return trip
while the rocket skims along the final past light-cone of the Milky Way.
The outward journey lasts only 9 Gyr or 11 Gyr in the proper time
measured by a comoving observer.  

The place where the rocket comes to rest finally depends quite
sensitively on the proper time when the rocket jerks.  When the rocket
goes from accelerating to decelerating, it is typically travelling
with $\beta\gamma \approx 1.4 \times 10^{10}$ (see Eq.~(\ref{eq:6}) or
even faster if $H$ decreases with time $w>-1$, so an error in the jerk
time of one second would result in the rocket coming to a stop about
140~pc off course.   In practice this error would not be terribly
important because by the time the rocket returned, the sun would have
long since become a white dwarf, and presumably the rocketeer would 
find little familiar upon her return.

\section{Conclusions}
\label{sec:conclusions}

The theoretical treatment of an accelerating rocket in a cosmlogical
setting is only marginally more difficult than in flat spacetime.  In
the de Sitter spacetime the rocket's trajectory may be solved
analytically.  However, unlike in flat spacetime, there is a limit to
the rapidity that a rocket can achieve in an expanding spacetime, so
after the rocket travels for a time of approximately $g^{-1}
\ln(g/H_0)$ or 23~years, the distance travelled by the rocket no longer
increases exponentially with the proper time of the rocketeer; in the
case of de Sitter space the distance increases linearly with the
proper time in this limit.

A more practical question is how much of the universe that we observe
today could we visit with future rocket technology.  The paths that
the rocket can complete in a finite proper time cannot reach all the
way to the future light cone of the Earth, but the rocket can go 99\%
of the way in a reasonable proper time of about a century;
consequently, if future technology can bring us to the stars,
travelling to distant galaxies could soon follow.

\bibliographystyle{prsty} 
\bibliography{mine,gr,cosmo}

\end{document}